\documentclass[conference]{IEEEtran}
\IEEEoverridecommandlockouts

\usepackage{cite}
\usepackage{amsmath,amssymb,amsfonts}
\usepackage{graphicx}
\usepackage{textcomp}
\usepackage{xcolor}
\usepackage{pgfplots}
\usepgfplotslibrary{fillbetween}
\pgfplotsset{compat=1.18}
\usepackage{booktabs} 
\usepackage{color,amsmath,amssymb,amsthm,epsfig,mathrsfs,cite,bm,graphics}
\usepackage{algpseudocode}
\usepackage{mathtools}
\usepackage{algorithm} 
\usepackage[utf8]{inputenc}
\usepackage{amsmath}

\usepackage{amsfonts}
\usepackage{amssymb}
\usepackage{placeins}
\usepackage{graphicx}
\usepackage{bbm,bm}
\usepackage{graphicx,graphics,subcaption}
\usepackage{tikz}
\usepackage{pgf}
\usepackage{pgfplots}
\usepackage[bookmarks=false]{hyperref}
\usepackage[all]{hypcap} 
\usepackage{float}
\usepackage{tabularx}
\usepackage{amsmath}

\usetikzlibrary{automata, positioning, arrows,shapes.multipart,fit,calc, angles,arrows.meta}
            
\usepackage{todonotes}
\presetkeys{todonotes}{color=blue!20}{}
\usepackage{pgfplots}
\usepackage{pgfplotstable}

\usepackage[letterpaper, left=0.625in, right=0.625in, bottom=1.03in, top=0.71in]{geometry}

\def\BibTeX{{\rm B\kern-.05em{\sc i\kern-.025em b}\kern-.08em
    T\kern-.1667em\lower.7ex\hbox{E}\kern-.125emX}}
\begin{document}

\title{RL-based Joint Coverage and Beam Optimization of High Altitude Platform Systems\\

\thanks{Guilhem Loussouarn, Nancy Nayak and Kin Leung were supported by the Dstl/RAF SDS Continuation Project, EP/Y037197/1 and EP/Y037243/1, respectively.}
}

\author{
\IEEEauthorblockN{Guilhem Loussouarn, Nancy Nayak and Kin K. Leung }
\IEEEauthorblockA{\textit{Department of Electrical and Electronic Engineering} \\
\textit{Imperial College London}, \\
London, SW7 2AZ,  UK\\
\{guilhem.loussouarn23,n.nayak,kin.leung\}@imperial.ac.uk
}
\and
\IEEEauthorblockN{Patrick J. Baker}
\IEEEauthorblockA{\textit{Rapid Capabilities Office} \\
\textit{Royal Air Force}\\
Farnborough, UK \\
pbaker@dstl.gov.uk}

}

\maketitle

\begin{abstract}
High Altitude Platform Systems (HAPS) are a promising component of 6G network architectures, offering a unique ``freedom of movement" that distinguishes them from static terrestrial networks (TN) and orbit-constrained satellite communications. This inherent mobility for HAPS provides a powerful mechanism to address non-stationarity, spatio-temporal user distributions, and traffic dynamics, such as periodic population migrations. This work addresses three key optimization problems in HAPS networks: (a) HAPS positioning for optimal coverage, (b) beam allocation, and (c) joint optimization of coverage and beam allocation. To tackle these complex challenges, a Reinforcement Learning (RL) framework is proposed, capable of operating in scenarios with multiple HAPS. The results demonstrate that the RL-based approach effectively learns to control HAPS positioning and resource allocation, dynamically adapting to variations in user distributions and traffic patterns. In particular, by employing a multi-policy Proximal Policy Optimization (PPO) approach, the proposed framework jointly learns HAPS positioning and allocating beams under spatio-temporal traffic demand variations and outperforms heuristic baselines. Simulation results demonstrate that our joint optimization approach significantly improves sum-rate and user satisfaction, showing that the dynamic mobility of HAPS can be successfully exploited to create highly responsive and efficient next-generation networks.
\end{abstract}


\begin{figure*}[ht]
\begin{equation}
\begin{aligned}
\label{eq:signalmodel}
y_{ihb}
&=
\mathbf{c}_i^T\sqrt{p_{h}}\mathbf{h}_{ih}^H \mathbf{w}_{hb} \sum_{i=1}^{N_{hb}} x_{ihb}\mathbf{c}_i +
\mathbf{c}_i^T\sum_{k\in\mathcal{B}_h\setminus b} 
\sqrt{p_{h}}\mathbf{h}_{ih}^H \mathbf{w}_{hk} \sum_{j=1}^{N_{hk}}x_{jhk}\mathbf{c}_j +
\mathbf{c}_i^T\sum_{\ell\in\mathcal{H}\setminus h}\sum_{p\in\mathcal{B}_\ell} 
\sqrt{p_{\ell}}\mathbf{h}_{i\ell}^H\mathbf{w}_{lp}  \sum_{g=1}^{N_{lp}}x_{glp} \mathbf{c}_g
+
\mathbf{c}_i^T n_{ihb}, 
\end{aligned}
\end{equation}
\vspace{-5mm}
\end{figure*}

\section{Introduction} 
Next-generation wireless networks face major challenges in achieving seamless connectivity using only conventional terrestrial infrastructure, given rapidly growing and dynamic user demand, as well as diverse application requirements \cite{lin2013towards, jang2026analytical}. Static TN suffers from coverage gaps, disaster vulnerability, and high deployment costs in remote regions. Satellite systems are constrained by bandwidth, signal degradation, high costs, and limited flexibility. Additionally, UAVs lack the endurance, mobility, and stability needed for large-scale, high-altitude coverage.

In this context, HAPS offer a promising complementary solution, providing flexible deployment, localized control, and dynamic adaptability due to their lower operating altitudes \cite{kurt2021vision}. This facilitates efficient enhancement of coverage and capacity while reducing reliance on rigid infrastructure and mitigating environmental challenges posed by satellites. A joint optimization framework for HAPS access links in \cite{javed2024system} addresses user grouping, association, beam steering, and power allocation to maximize sum rate and system efficiency. The feasibility of large-scale HAPS deployment for island and maritime coverage, considering realistic shadowing and channel models, is studied in \cite{lin2026haps}, where HAPS mainly serve island and maritime scenarios. Coverage probability and ergodic capacity for a HAPS on a fixed circular trajectory are analyzed in \cite{jang2026analytical}, but the static network assumptions limit its application to dynamic settings. To the best of our knowledge, coverage optimization for multiple mobile HAPS with adaptive trajectories under varying user distributions remains unaddressed.

In this work, we address both coverage optimization and user throughput within session duration, aiming for low-latency communication. An integrated network with multiple HAPS and ground base stations is studied in \cite{liu2026sum}, which jointly optimizes throughput and fairness via user association and beamforming using classical optimization techniques, such as a generalized assignment problem approach and successive convex approximation. However, \cite{liu2026sum} assumes static traffic and fixed HAPS positions across different regions, overlooking the need for dynamic repositioning and incurring high computational complexity from repeated optimization.

In \cite{srinivasan2021airplane}, an air-to-ground backbone integrates space and TNs by employing HAPS as mobile base stations with millimeter-wave communication. The study utilizes three beamforming techniques based on large-scale planar arrays to enhance directional gain and suppress inter-user interference. However, at typical HAPS altitudes, user-specific beamforming is impractical due to excessive antenna requirements and payload constraints. To address this, we employ a beam-hopping (BH) strategy, where fewer antennas serve groups of co-located users using distinct codes \cite{javed2024system}. In dynamic environments with varying user density and traffic, repeated conventional optimization is computationally intensive and unsuitable for real-time operation. Instead, we utilize reinforcement learning (RL) to dynamically control fully mobile HAPS, adapting network topology to real-time user demand. This approach is particularly effective during localized surges, such as large events, enabling HAPS to adjust positions and coverage to balance loads and maintain efficient service continuity.

In summary, we propose an RL-based approach to determine near-optimal HAPS positioning for maximizing coverage under varying user distributions and traffic. We further introduce an RL-based BH strategy for multiple static HAPS, explicitly modeling dynamic user arrivals and departures, unlike existing methods with fixed user sets. Finally, we develop an RL framework for joint HAPS positioning and BH sequence learning in scenarios with periodically changing user distributions and multiple mobile HAPS. Simulation results demonstrate that our approach outperforms benchmarks by serving more users and significantly improving average demand satisfaction and system throughput.
\section{System Model}

Consider a set $\mathcal{H}$ with $|\mathcal{H}|=H$ of HAPS that serve ground users distributed over a geographical area. Each HAPS is equipped with a uniform rectangular array (URA) composed of $A_{x}\times A_{y}$ transmit antennas, while users are equipped with a single omnidirectional receiving antenna. The service area of each HAPS is divided into $M$ subzones such that each subzone can be served by a single beam. The users in the same subzone are served using different codes to avoid inter-user interference. At each time step $t$, HAPS $h$ generates a set of beams $\mathcal{B}_h^t$ with $|\mathcal{B}_h^t|=B$. Each HAPS $h$ forms $B \leq M$ beams, each directed towards a different subzone and characterized by transmit beamformer $\mathbf{w}_{hb}$. All users in subzone $b$ receive the same beam $\mathbf{w}_{hb}$. However, beams for different subzones and HAPS may cause mutual interference.

Let $N_{hm}^t$ denote the number of users in subzone $m$ of HAPS $h$ at time $t$, and $N_{hb}^t$ the number served by beam $b$ of HAPS $h$. The signal transmitted from HAPS $h$ for user $i$ in subzone $b$ is $x_{ihb}$. Users in the same subzone are assigned Walsh-Hadamard codes, where user $i$ is assigned with $\mathbf{c}_i$ and
\begin{equation}
    |\mathbf{c}_i|^2 = 1, \text{ and } \mathbf{c}_i^T \mathbf{c}_k =
\begin{cases}
1, & \text{if } i = k\\
0, & \text{if } i \neq k.
\end{cases}
\end{equation}
The signal for subzone $b$ of HAPS $h$ is then $x_{hb} = \sum_{i=1}^{N_{hb}^t} x_{ihb}\mathbf{c}_i$. The signal received at the user $i$ is given by \eqref{eq:signalmodel}, where $p_{h}$ is the transmit power of HAPS $h$, and $n_{ihb} \sim\mathcal{CN}(0,\sigma^2)$.
The transmit beam steering vector from HAPS $h$ to subzone $m$ at time step $t$ is given by 
\begin{equation}
\begin{aligned}
&\mathbf{w}_{hm}^t = \mathbf{a}(\theta_{hm},\phi_{hm})= \mathbf{a}_{N_{x}^{}}^{}(\theta_{hm}^{},\phi_{hm}^{})\otimes \mathbf{a}_{N_{y}^{}}^{}(\theta_{hm}^{}, \phi_{hm}^{}) \\
&\mathbf{a}_{A_{x}^{}}^{}(\theta_{hm}^{},\phi_{hm}^{}) = [1, \dots, e^{j2\pi \frac{d_a}{\lambda} A_{x}^{} \cos{\phi_{hm}^{}} \sin{\theta_{hm}^{}}}, \dots, \\& e^{j2\pi \frac{d_a}{\lambda} (A_{x}^{}-1) \cos{\phi_{hm}^{}}\sin{\theta_{hm}^{}}}]^H\\
&\mathbf{a}_{A_{y}^{}}^{}(\theta_{hm}^{}, \phi_{hm}^{})  = [1, \dots, e^{j2\pi \frac{d_a}{\lambda} a_{y}^{} \sin{\phi_{hm}^{}}\sin{\theta_{hm}^{}}}, \dots, \\& e^{j2\pi \frac{d_a}{\lambda} (A_{y}^{}-1) \sin{\phi_{hm}^{}}\sin{\theta_{hm}^{}}}]^H,
\end{aligned}
\end{equation}
where $\theta_{hm}$, and $\phi_{hm}$ are elevation and azimuth angles, respectively, and $d_a$ is antenna element separation \cite{balanis2015antenna}. To serve $B$ geographical subzones, each HAPS selects beam directions from a predefined angular codebook comprised of $\mathbf{w}_{hm}^t$ where $\theta_{hm}$, and $\phi_{hm}$ $\forall h\in \mathcal{H}$ are discretized as
\begin{align}
\theta_{hm}\in \{ l\Delta_\theta \mid l\in\mathcal{K}_\theta \},
\phi_{hm} \in \{ k\Delta_\phi \mid k\in\mathcal{K}_\phi \},
\end{align}
where $\Delta_\phi$ and $\Delta_\theta$ are the angular resolutions. Due to strong line-of-sight, the channel between user $i$ and HAPS $h$ is modelled as Rician with $K_{ih}$ as Rician factor \cite{lin2026haps}, 
Here, $\beta_{ih}
=
G_t G_r
10^{-\mathrm{PL}_{avg}(d_{ih},\tilde{\theta}_{ih})/10}
A_{atm}(f,d_{ih})$ is the large-scale channel gain, where $\mathrm{PL}_{avg}(d,\tilde{\theta})$ denotes the average path loss and $A_{atm}(f,d)$ models atmospheric attenuation \cite{Grace2010}. Here $d_{ih} = \sqrt{h_z^2 + h_{xy}^2}$ is the slant distance, $h_{xy}$ is the horizontal distance, and $\tilde{\theta}_{ih} = \arctan\!\left(\frac{h_z}{h_{xy}}\right)$ is the elevation angle between user $i$ and HAPS $h$. 
Based on \eqref{eq:signalmodel}, the signal-to-interference-plus-noise ratio (SINR) experienced by user $i$ in subzone $b$ of HAPS $h$ at time step $t$ is given by
\begin{equation}
\begin{aligned}
\label{eq:sinr}
    \gamma_{ih}^t =
\frac{p_{h}|\mathbf{h}_{ih}^H\mathbf{w}_{hb}|^2}{
\sum\limits_{k\in\mathcal{B}_h\setminus b}^{}
p_{h}|\mathbf{h}_{ih}^H\mathbf{w}_{hk}|^2
+
\sum\limits_{\ell\in \mathcal{H}\setminus h}\sum\limits_{p\in\mathcal{B}_\ell}
p_{\ell}|\mathbf{h}_{i\ell}^H\mathbf{w}_{\ell p}|^2
+
\sigma^2 },
\end{aligned}
\end{equation}
resulting in a data rate of $r_{ih}^t= \log_2\left(1 + \gamma_{ih}^t\right).$ With time dependence, $\gamma_{ih}^t$, $r_{ih}^t$, and $\mathbf{w}_{hm}^t$ denote SINR, data rate, and beam directions at time $t$.

The service area covered by $H$ HAPS is divided into $U$ sub-areas to capture spatial demand variations. User arrivals in sub-area $u$ follow a Poisson process with rate $\lambda_u(t)$, and session durations are modeled as Geometric with parameter $\mu_u(t)$. The set $\mathcal{U}^t$ contains all users present at time $t$, reflecting the underlying arrival and departure processes. For fine-grained allocation (e.g., BH), user arrivals and departures are time-invariant, $\lambda_u(t) = \lambda_u$, $\mu_u(t) = \mu_u$, named as \textit{stationary environment}. For long-term objectives, a \textit{periodic environment} is assumed with $\lambda_u(t + T) = \lambda_u(t)$ and $\mu_u(t + T) = \mu_u(t)$ to model temporal changes in demand.

The three optimization objectives address different aspects of system design. The first objective is to optimize the \textit{positions of the HAPS} to maximize the total number of users connected to HAPS $h$, denoted by $C_h^t$, while minimizing the $I^t$, the number of users among the total connected users experiencing overlapping coverage. The objective is 
\begin{align}
\label{eq:CovObj}
    \max_{\{\{ x_h^t, y_h^t, z_h^t\}_{h = 1}^H\}_{t=1}^{\tau_1}}\sum_{t=1}^{\tau_1} \sum_{h = 1}^H  C_h^t - I^t
\end{align}
where $(x_h^t, y_h^t, z_h^t)$ denote the coordinates of HAPS $h$. The objective is to find the optimal HAPS positions over a window $\tau_1$ to maximize served users and minimize inter-platform interference by avoiding overlapping coverage.

Coverage alone does not fully capture system performance. Thus, the second objective optimizes BH sequences to maximize throughput and user satisfaction by dynamically allocating beams across space and time to serve diverse user demands. As user and traffic variations are slower than BH decisions, a stationary environment is assumed. Each user $i$ is active from $T_i^{start}$ to $T_i^{end}$, with a session duration $T_i=T_i^{end}-T_i^{start}$ assumed to be reported by users. When computing cumulative data rate up to time $t$, $r_{ih}^\tau$ is set to zero if user $i$ is not served at time steps $\tau \leq t$. The cumulative data rate for user $i$ is $R_i^t = \sum_{\tau=1}^{t} r_{ih}^{\tau}$. Let each user request a target rate $\mu^{\text{rate}}$, so the target cumulative data for user $i$ is $\mu_i = \mu^{\text{rate}} T_i$. We define the completion ratio $c_i^t = \min\left(1, \frac{R_i^t}{\mu_i}\right)$ for user $i$ at time step $t$, which quantifies the extent to which the user's target cumulative data rate has been satisfied by the BH mechanism up to time $t$. $c_i^t=1$ when the cumulative data rate $R_i^t$ attained by user $i$ up to time step $t$ is equal to the user's target requirement $\mu_i$. The goal is to optimize the transmit beamformers associated with the BH scheme, with the overall objective function presented as
\begin{equation}
\begin{aligned}
\label{eq:BHobjective}
    \max_{\{\{\{\mathbf{w}_{hb}^t\}_{b \in \mathcal{B}_h^t}\}_{h=1}^{H}\}_{t=1}^{\tau_2}} \quad & \sum_{t=1}^{\tau_2} \left[\frac{1}{|\mathcal{U}^t|} \left(w_1.\bar{r}^t + w_2. \bar{c}^t + w_3. \bar{s}^t  \right)\right] \\
\text{s.t. } \sum_{b \in \mathcal{B}_h^t} \|\mathbf{w}_{hb}^t\|^2 \le p_h, & \text{ where }
\bar{r}^t = \sum_{h\in\mathcal{H}} \sum_{b=\mathcal{B}_h^t}\sum_{i=1}^{N_{hb}^t} r_{ih}^t, \\ \bar{c}^t = \sum_{h\in\mathcal{H}} \sum_{m=1}^{M}\sum_{i=1}^{N_{hm}^t} c_i^t, &\text{ and }\bar{s}^t = \sum_{h\in\mathcal{H}} \sum_{m=1}^{M}\sum_{i=1}^{N_{hm}^t} \mathbb{I}\{c_i^t = 1\}
\end{aligned}
\end{equation}
Here, the term $\bar{r}^t$ represents the \textit{aggregate system sum rate} corresponding to the served users by the $B$ beams of each HAPS. The second term $\bar{c}^t$ is \textit{aggregate completion} and is defined as the average degree to which the users' target throughput requirements are satisfied by the HAPS. The third term $\bar{s}^t$ is the \textit{aggregate success} which promotes the rapid satisfaction of users' target throughput requirements. Since the number of users changes at every time step, the weighted sum of $\bar{r}^t$, $\bar{c}^t$, and $\bar{s}^t$ is further normalized by $|\mathcal{U}^t|$.

Optimizing coverage and BH sequences separately is insufficient; thus, the third objective jointly optimizes HAPS positioning and BH sequence prediction in a unified framework. The optimization objective is given by
\begin{equation}
\begin{aligned}
\label{eq:Jointobjective}
    \max_{\substack{
\{\{x_h^t, y_h^t, z_h^t, \\\{\mathbf{w}_{hb}^t\}_{b \in \mathcal{B}_h^t}\}_{h=1}^{H}  
\}_{t=1}^{\tau_3}
}} & \sum_{t=1}^{\tau_3} \left[\frac{1}{|\mathcal{U}^t|} \left(w_1.\bar{r}^t + w_2. \bar{c}^t + w_3.\bar{s}^t  \right)\right] \\
 \text{s.t.} \sum_{b \in \mathcal{B}_h^t} \|\mathbf{w}_{hb}^t\|^2 &\le p_h, \text{ with } \bar{r}^t, \bar{c}^t, \bar{s}^t \text{same as }\eqref{eq:BHobjective}.
\end{aligned}
\end{equation}
Notably, the joint optimization retains the BH objective, but expands the decision variables to include both HAPS positions and beam allocation.
This integrated approach leverages the interdependence between platform positioning and dynamic beam allocation, resulting in improved network coverage and capacity compared to separate optimization. The metrics $\bar{r}^t, \bar{c}^t, $ and $\bar{s}^t$ are linearly related to the number of connected users and are influenced by interference. To solve the objectives in \eqref{eq:CovObj}, \eqref{eq:BHobjective}, and \eqref{eq:Jointobjective}, we propose learning-based techniques to enable adaptive resource management in response to dynamic network conditions.

\section{Learning Methods}
The above optimization problems can be modeled as Markov Decision Processes (MDPs). In the case of a stationary environment, an MDP is defined by a tuple $\left( \mathcal{S},\mathcal{A}, P, R, \gamma \right)$, with $P$ being the transition dynamics in the stationary environment. Let $\mathcal{S}$ denote the state space, $\mathcal{A}$ the action space, and $R$ the reward associated with a state-action pair. In contrast, a periodic environment is modeled by a periodic MDP (pMDP) defined by the tuple $\left( \mathcal{S},\mathcal{A}, P(k), R, \gamma, K \right)$ with $K$ being the number of different phases in a period and $P(k)$ being the transition probabilities.
In this work, an actor-critic, policy-gradient RL algorithm, namely PPO \cite{SchulmanWDRK17}, is employed to address the high-dimensional and discrete nature of the action spaces. The policy is parameterized by $\psi$ and comprises a lightweight feature-extraction stage followed by separate actor and critic networks. The input observations are first processed using a flattening operation, after which the actor head outputs the policy $\pi_{\psi}(a^t|s^t)$, and the critic head estimates the state-value function $V_{\psi}(s^t)$. Both the actor and critic are implemented as feedforward neural networks with identical architectures but independent parameters. The training objective is to minimize a loss function $L^t(\psi)$ where
\begin{equation}
\label{eq:ppo}
L^t(\psi) = \hat{\mathbb{E}}_t \left[ L_{\mathrm{CLIP}}^t(\psi) - c_1 L_{\mathrm{VF}}^t(\psi) + c_2 S[\pi_{\psi}](s^t) \right],
\end{equation}
where $c_1$ and $c_2$ are coefficients that weight the value function error and the entropy bonus, respectively \cite{SchulmanWDRK17}. The policy surrogate is defined as, $L_{\mathrm{CLIP}}^t(\psi) = \hat{\mathbb{E}}_t \left[ \min\left(r^t(\psi) \hat{A}^t,\ \text{clip}\left(r^t(\psi), 1 - \epsilon, 1 + \epsilon\right)\hat{A}^t\right) \right],$ where $r^t(\psi) = \frac{\pi_{\psi}(a^t|s^t)}{\pi_{\psi_{\text{old}}}(a^t|s^t)}$ is the probability ratio between the current and previous policies, and $\hat{A}^t = R^t + \gamma V_{\psi}(s^{t+1}) - V_{\psi}(s^t)$. 
The return is computed backward from the end of the episode of length $\tau_1$, starting with $\hat{R}^{\tau_1} = R^{\tau_1}$ and recursively as $\hat{R}^t = R^t + \gamma \hat{R}^{t+1}$ for $t = \tau_1 - 1, \ldots, 1$. The value function error is defined as $L_{\mathrm{VF}}^t(\psi) = \hat{\mathbb{E}}_t \left[ (V_{\psi}(s^t) - \hat{R}^t)^2 \right]$, ensuring better evaluation of the states by the critic. The entropy bonus is expressed in such a way that if the agent has $n$ possible discrete actions $S[\pi_{\psi}](s^t) = -\sum_{i=1}^{n} \pi_{\psi}(a_i|s^t) \log \pi_{\psi}(a_i|s^t)$. By subtracting the entropy in the loss, the optimization will try to increase the entropy and avoid falling into one deterministic action. 

To solve \eqref{eq:CovObj}, we adopt a centralised RL approach based on PPO where the state $s^t\in\mathcal{S}$ is defined as:
\begin{equation}
\label{eq:stateCov}
    s^t = \left[ \{x_h^t, y_h^t, z_h^t, \frac{C_h^t}{|\mathcal{U}^t|}\}_{h=1}^H ,\frac{I^t}{|\mathcal{U}^t|}\right].
\end{equation}
The reward at each time step is defined as
\begin{equation}
\label{eq:rewardCov}
    R^t = \sum_{h=1}^H  C_h^t - I^t
\end{equation}
and the action $a^t\in\mathcal{A}$ consists of discrete displacements $\Delta_x^h, \Delta_y^h, \Delta_z^h \in \{-1,0,1\}$, given by
\begin{equation}
\label{eq:actionCov}
    a^t = [\{ \Delta_x^h, \Delta_y^h, \Delta_z^h \}_{h=1}^H], 
\end{equation}
where $x_h^{t+1} = x_h^t + \Delta_x^h d_x$. We first evaluate the agent in a stationary setting. However, real-world traffic follows periodic patterns. Therefore, we consider a non-stationary but periodic environment. Let the episode of length $\tau_1$ considered for the PPO algorithm consist of $K$ periods of duration $\tau$ time steps each, resulting in a total duration of the episode of $\tau_1= K\tau$ time steps. To address this periodic non-stationarity, we propose two methods: a single augmented policy and a multi-policy approach. 

\textit{Single-policy PPO (sPPO-C):} Learning a single policy for a non-stationary environment is challenging for an RL agent \cite{chen2022adaptive}. However, for a periodic non-stationary environment, we propose to augment a phase indicator $\varphi^t = \left\lfloor \frac{t}{\tau} \right\rfloor \bmod K$ in $s^t$ such that $s^t = \left[ \{x_h^t, y_h^t, z_h^t, \frac{C_h^t}{|\mathcal{U}^t|}\}_{h=1}^H,\frac{I^t}{|\mathcal{U}^t|},\varphi^t\right]$, which enables to learn a single policy for a periodic environment. The reward $r^t$ and action $a^t$ remain identical to \eqref{eq:rewardCov} and \eqref{eq:actionCov}, respectively.

\textit{Multi-policy PPO (mPPO-C)}: Unlike a single-policy PPO agent that must learn a complex mapping for all time steps in the episode, the multi-policy PPO algorithm learns $K$ specialized policies given by $\psi_k,\,\forall k\in\{1,\dots,K\}$, optimized for the $k^{th}$ phase of the traffic cycle. This significantly reduces the learning variance caused by periodic distribution shifts, and policies can be designed for a non-stationary environment by exploiting the periodic nature of traffic evolution \cite{CHEN2025111645}. 
 
The states, reward, and actions are the same as described in equations \eqref{eq:stateCov}, \eqref{eq:rewardCov}, and \eqref{eq:actionCov}, respectively.

BH demonstrated in \eqref{eq:BHobjective} is addressed by formulating an MDP and then appropriately modifying the PPO algorithm to maximize the combined objectives of throughput and user satisfaction, and is denoted by \textit{PPO-BH}. For fine-grained resource allocation such as BH, the traffic demands are assumed to be stationary, and the HAPS are maintained at fixed locations during BH operations. The HAPS autonomously select beams from a predefined codebook, dynamically adjusting allocations to accommodate varying user requirements. At each time step $t$, the state $s^t$ combines individual HAPS-wise features with global system-level information, capturing recent and historical states to comprehensively represent active users and traffic patterns.

\textit{Per-HAPS features}: For each HAPS $h$, key features are constructed to capture user distribution and BH history for scheduling decisions. The \textit{visible user fraction} is defined as the ratio of users observed within the coverage footprint of HAPS $h$ and the total users at time step $t$, given by $f_{\mathrm{vis},h}^{t} = 1/|\mathcal{U}^{t}|.\sum_{m=1}^M N_{hm}^t$. The \textit{served user fraction} $f_{\mathrm{srv},h}^{t} = 1/|\mathcal{U}^{t}|.\sum_{b \in \mathcal{B}_h^t}N_{hb}^t$ reflects the proportion of users currently being served by the set of beams $\mathcal{B}_h^t$ assigned to HAPS $h$. To represent the spatial user distribution, we employ an \textit{angular user density map} $\mathbf{D}_h^t$ where the discretized measure of user concentration per subzone $m$ is provided by $D_{hm}^{t} = 1/|\mathcal{U}^{t}|.N_{hm}^t.$ A \textit{beam visit history map} $\boldsymbol{V}_h^{t}$ is formed where element $m$ tracks the temporal visit frequency of subzone $m$ of HAPS $h$, and $V_{hm}^{t} = \alpha V_{hm}^{t-1} + \mathbb{I}\{m \in \mathcal{B}_h^t\},$ where $\alpha \in [0,1)$ is a forgetting factor and $\mathbb{I}\{\cdot\}$ is the indicator function. Finally, a \textit{completion deficit map} $\boldsymbol{F}_h^{t}$ is introduced, with each subzone $m$ containing $F_{hm}^{t} = 1/N_{hm}^t. \sum_{i =1}^{N_{hm}^t} (1 - c_i^{t}).$

\textit{Global features}: In addition to per-HAPS observations, several global system-level features are included to comprehensively describe network status at each time step. For example, $|\mathcal{U}^{t}|$ represents the total user load, and $\bar{r}^t$ quantifies the overall throughput achieved across all HAPS and beams. Furthermore, user completion statistics, e.g., $\bar{c}^t$ and $\bar{s}^t$, are included to assess progress toward user-specific data delivery targets. Finally, a time encoding component $(\sin(2\pi t/\tau_2), \cos(2\pi t/\tau_2))$ is appended to provide the learning agent with periodic temporal context, enabling it to capture and exploit any inherent time-dependent structure in the traffic demand. Finally the state is given by $s^t =  [\{f_{\mathrm{vis,h}}^{t}, f_{\mathrm{srv},h}^{t}, \mathbf{D}_h^{t}, \mathbf{V}_h^{t}, \mathbf{F}_h^{t}\}_{h=1}^H, |\mathcal{U}^t|, \frac{\bar{r}^{t}}{|\mathcal{U}^t|}, \frac{\bar{c}^t}{|\mathcal{U}^t|}, \frac{\bar{s}^t}{|\mathcal{U}^t|},\\ \sin(2\pi t/\tau_2), \cos(2\pi t/\tau_2)]$. The reward $R^t$ is given by: 
\begin{equation}
\label{eq:rewardBH}
    R^t = \left( w_1 \cdot \bar{r}^t + w_2 \cdot \bar{c}^t + w_3 \cdot \bar{s}^t \right)/|\mathcal{U}^t|
\end{equation}

Each HAPS selects $B$ distinct beams from a predefined codebook to form its transmission pattern at each time step. To manage combinatorial complexity, all possible beam selection combinations are enumerated and indexed, resulting in a discrete action set $\{\{\mathbf{w}_{hb}\}_{b \in \mathcal{B}_h}\}_{h=1}^H$. The total number of feasible actions is $\binom{M}{B}^H$. 
To measure the performance of the proposed methods, we report average system sum rate $\bar{r}_{\text{avg}}=1/\tau_2\sum_{t=1}^{\tau_2}\bar{r}^t$, average reward $R_{\text{avg}}=1/\tau_2\sum_{t=1}^{\tau_2}R^t$, average success in the system at time step $t$ $\bar{s}_{\text{avg}}=1/\tau_2\sum_{t=1}^{\tau_2}\frac{\bar{s}^t}{|\mathcal{U}^t|}$, average completion with respect to the users that are served by the HAPS $\bar{c}_{\text{avg}} = 1/\tau_2\sum_{t=1}^{\tau_2} \frac{\bar{c}^t}{\sum_{h\in \mathcal{H}}\sum_{m=1}^M N_{hm}^t}$, and average served user fraction $\bar{f}_{\text{srv,avg}}= 1/\tau_2\sum_{t=1}^{\tau_2}\sum_{h\in\mathcal{H}}f_{\text{srv,h}}^t$ for an episode length of $\tau_2$.

\begin{algorithm}[!t] 
\caption{Proposed mPPO-JCBH}
\label{alg:multi_ppo}
\begin{algorithmic}[1]
\State Initialize $\psi_1,\dots,\psi_K$ and buffers $\mathcal{B}_1, \dots,\mathcal{B}_K$
\For{episode $=1,\dots,E$}
    \For{$t=1,\dots,\tau$}
        \If{$t \in$ phase $k$}
            \State Sample $a^t \sim \pi_{\psi_k}(s^t)$, observe $(R^t,s^{t+1})$
            \State Store $(s^t,a^t,R^t,\log\pi_{\psi_k},V_{\psi_k})$ in $\mathcal{B}_k$
        \EndIf
    \EndFor
    \For{$k=1,\dots,K$}
        \State Compute $\hat{A}, \hat{R}$ from $\mathcal{B}_k$
        \For{epoch $=1,\dots,N$}
            \State Sample batch from $\mathcal{B}_k$, update $\psi_k$ using \eqref{eq:ppo}
        \EndFor
        \State Clear $\mathcal{B}_k$
    \EndFor
\EndFor
\end{algorithmic}
\end{algorithm}

For joint optimization of HAPS positioning and BH (JCBH) presented in \eqref{eq:Jointobjective}, the formulated MDP has the action $a^t = \left[\{\Delta_x^h, \Delta_y^h, \Delta_z^h, \mathbf{w}_{hb}\}_{b \in \mathcal{B}_h}\}_{h=1}^H \right]$. The state $s^t$ includes the position of the HAPS and connected users such that,
\begin{equation}
\begin{aligned}
    \label{eq:state_Cov_BH}
    s^t &= [ \{  x_h^t, y_h^t, z_h^t, C_h^t/|\mathcal{U}^t|, f_{\mathrm{vis,h}}^{t},f_{\mathrm{srv},h}^{t},\mathbf{D}_h^{t},\mathbf{V}_h^{t}, \mathbf{F}_h^{t}\}_{h=1}^H,\\&|\mathcal{U}^t|,\bar{r}^{t}/|\mathcal{U}^t|,\bar{c}^t/|\mathcal{U}^t|,\bar{s}^t/|\mathcal{U}^t|,\sin(2\pi t/\tau_3), \cos(2\pi t/\tau_3)].
\end{aligned}
\end{equation}
The number of users experiencing coverage from multiple HAPS, denoted by $I^t$ used in \eqref{eq:stateCov}, is not required in \eqref{eq:state_Cov_BH}, as it is already implicitly present in $\bar{r}^t$. The reward used is the same as in \eqref{eq:rewardBH}. To solve the MDP associated with JCBH in a stationary environment, a PPO is used and is denoted as \textit{PPO-JCBH}, whereas to solve the same MDP for a non-stationary periodic environment, the multi-policy PPO algorithm, denoted by mPPO-JCBH, is proposed, which is outlined in Alg. \ref{alg:multi_ppo}.

\section{Results}
\label{sec:simulationresults}
In this section, we provide simulation results for serving an area of $300 \times 300$ km$^2$ and an altitude between $18$ km and $20$ km. The geographical area has been equally divided into $9$ squares. The maximum allowable HAPS speed is $d_x=1$ km per time step. For the stationary environment, we have considered an episode of length $500$ time steps, whereas for periodic environments, we consider an episode of length $3000$ time steps, incorporating $2$ periods where each period has $K=3$ phases, with each phase being $500$ time steps. For the beamforming configuration, each HAPS can serve up to $2$ sub-zones simultaneously with $B = 2$ beams, employing beams selected from a codebook with azimuthal resolution of $60^\circ$ and elevation resolution of $30^\circ$, and a total number of possible beam directions of $M = 12$. We assume a carrier frequency of $f_c=3.5$ GHz with $20$ MHz bandwidth. The HAPS transmit power is $10$ W, and the noise power is $4 \times 10^{-13}$ W. Each HAPS employs a $4 \times 4$ URA with element spacing of $3e8/f_c$, $\mu^{\text{rate}}=0.5$, $(w_1,w_2, w_3) = (50, 15, 50)$, $K_{ih}=10$ dB, and $\bar{r}_{\text{avg}}$ is in bits/sec/Hz. 
The traffic model employs a spatially heterogeneous arrival rate $\lambda_u \in[0.01,3.0]$ and a geometric departure probability $\mu_u \in [0.05, 0.2]$, resulting in average user lifetimes of $5$ to $20$ time steps. We vary $\lambda_u$ and $\mu_u$ to generate diverse density zones, including hotspots. To simulate spatio-temporal dynamics in the environment, $\lambda_u$ is adjusted across different phases of the periods in non-stationary periodic environments. The developed PPO algorithms use a learning rate of $10^{-4}$ and an entropy coefficient of $0.01$. Results are obtained using a CPU-based standard laptop with $16$ GB RAM.

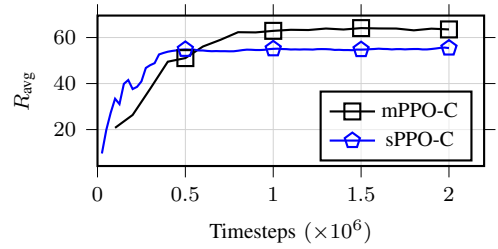
\begin{figure}[!t]
\centering
\begin{tikzpicture}
\begin{axis}[
    width=0.75\columnwidth,
    height=0.4\columnwidth,
    xlabel={Timesteps $(\times10^6)$},
    ylabel={$R_{\text{avg}}$},
    xmin=0,
    axis line style={black},
    tick align=outside,
    tick style={black},
    line width=0.8pt,
    grid=both,
    grid style={draw=gray!15, line width=0.2pt},
    major grid style={draw=gray!35, line width=0.3pt},
    tick label style={font=\footnotesize},
    label style={font=\footnotesize},
    legend pos=south east,
    legend style={fill opacity=0.6, draw opacity=1.0, text opacity=1.0, font=\footnotesize},
    scaled ticks=true
]

\addplot[
    black,
    solid,
    thick,
    mark=square,
    mark size=3.0,
    mark repeat=5,
    mark phase=5
]
table[
    col sep=comma,
    x=timesteps,
    y=mean_raw_reward_per_step
]{data_HAP/raw_eval_curve_Periodic_Multipolicy.csv};
\addlegendentry{mPPO-C}

\addplot[
    blue,
    solid,
    thick,
    mark=pentagon,
    mark size=3.0,
    mark repeat=20,
    mark phase=20
]
table[
    col sep=comma,
    x=timesteps,
    y=mean_raw_reward_per_step
]{data_HAP/raw_eval_curve_Periodic_Single.csv};
\addlegendentry{sPPO-C}

\end{axis}
\end{tikzpicture}
\caption{Evolution of learning for coverage optimization.}
\label{fig:periodic_comparison_ieee}
\vspace{-3mm}
\end{figure}

\begin{figure}[ht] 
     \centering
  
     \begin{subfigure}[b]{0.18\textwidth}
         \centering
         \includegraphics[width=\textwidth]{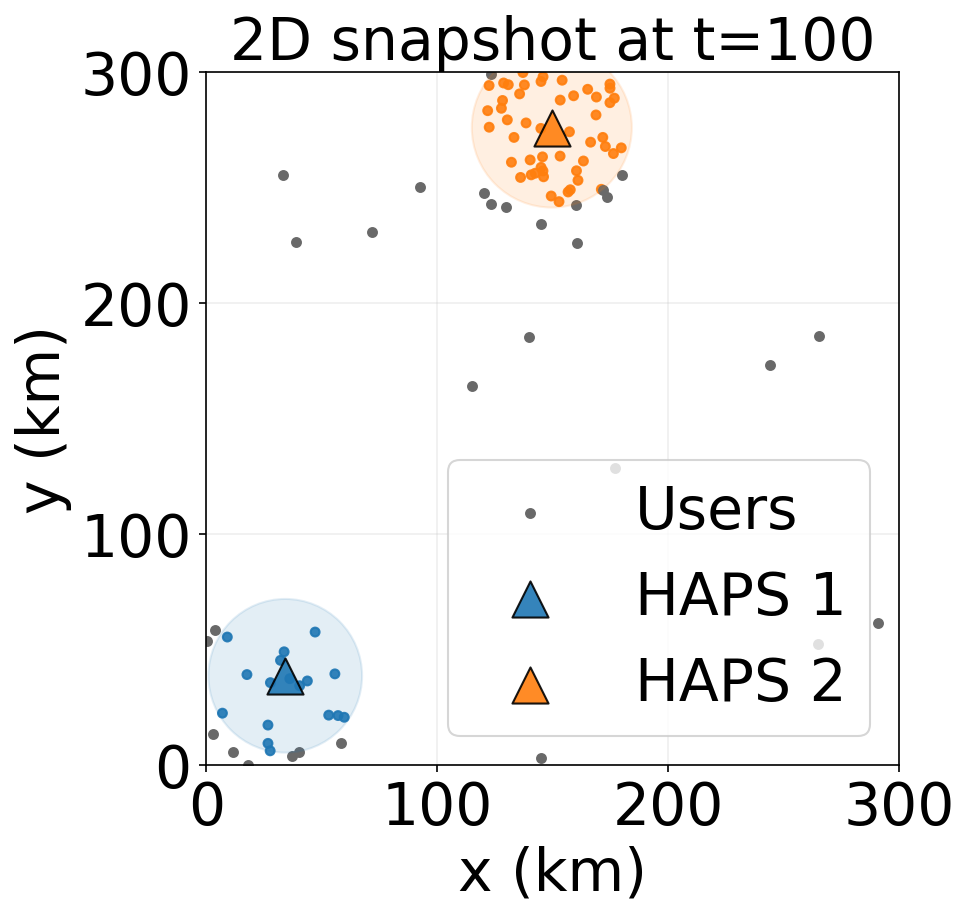}
     \end{subfigure}%
     \begin{subfigure}[b]{0.18\textwidth}
         \centering
         \includegraphics[width=\textwidth]{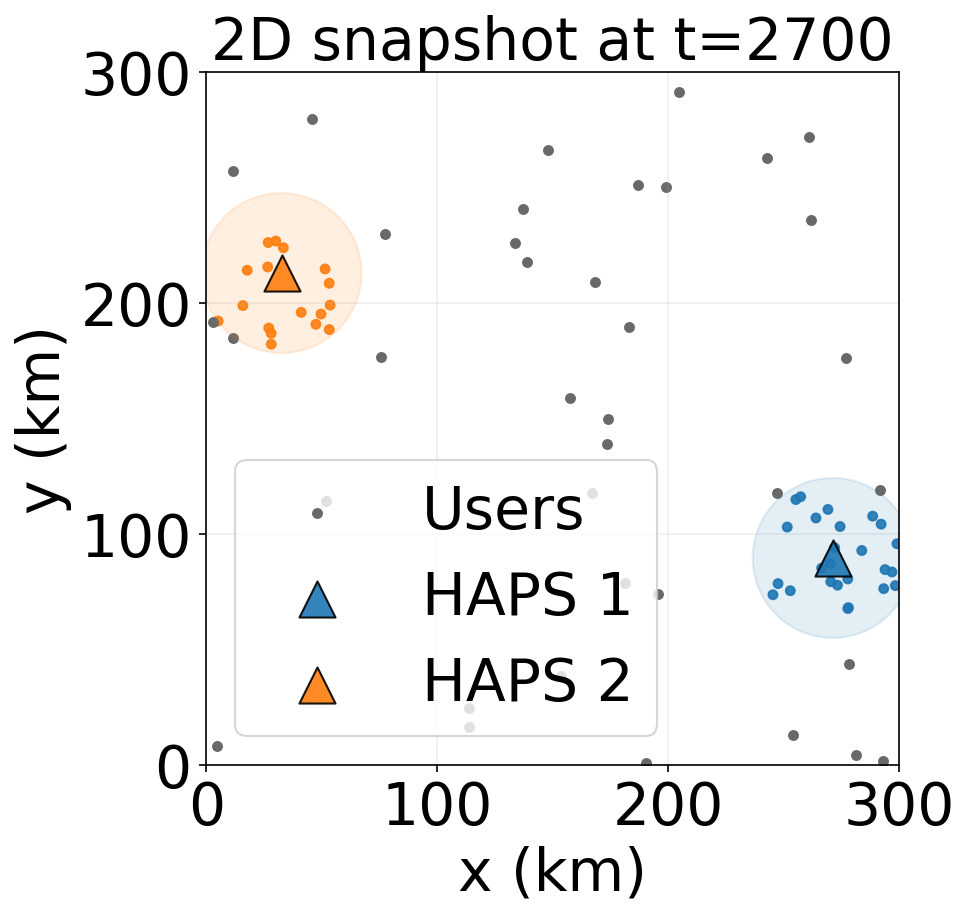}
     \end{subfigure}

     \caption{HAPS positioning ($H=2$) by mPPO-C at phases $k=1$ (left) and $k=3$ (right) as examples.}
     \label{fig:Pos1}
     \vspace{-3mm}
\end{figure}

\begin{table}[ht]
\centering
    \caption{Comparisons of BH policies }
\label{tab:policy_comparison_BH}
\begin{tabular}{lccccc}
\toprule
Policy & $R_{\text{avg}}$ & $\bar{r}_{\text{avg}}$  &   $\bar{s}_{\text{avg}}$ &$\bar{c}_{\text{avg}}$ & $\bar{f}_{\text{srv,avg}}$ \\
\midrule
PPO-BH & 27.21 & 12.47 & 20.07\% & 95.02\% & 15.25\%\\
Random beam (RB) & 9.92 & 3.81 & 6.26\% & 66.09\% & 4.65\%  \\
Beam sweeping (BS) & 9.95 & 4.26 & 8.63\% & 69.66\% & 5.09\%  \\
\bottomrule
\end{tabular}
\label{tab: AIBH}
\vspace{-3mm}
\end{table}

For coverage optimization, it is relevant to investigate the applicability of the proposed solutions in response to the spatio-temporal variations in user density in the environment. The following results are presented for a periodic environment with $K=4$ phases; however, the findings remain applicable under stationary conditions. In the case of a periodic environment, both sPPO-C and mPPO-C algorithms are employed. From the evolution of learning over training duration presented in Fig. \ref{fig:periodic_comparison_ieee}, we see that mPPO-C converges more slowly to a solution compared to sPPO-C, but mPPO-C converges to a solution with $15\%$ higher reward compared to sPPO-C. Fig. \ref{fig:Pos1} presents the visualisation of HAPS positioning at different phases by the mPPO-C algorithm for $H=2$. It can be seen that as the user distribution changes across phases, and the locations of the two hotspots vary in each phase, the mPPO-C solution effectively positions the two HAPS above the respective hotspots at each step. Given the demonstrated effectiveness of the multipolicy approach in periodic environments, subsequent experiments employ multi-policy algorithms. 
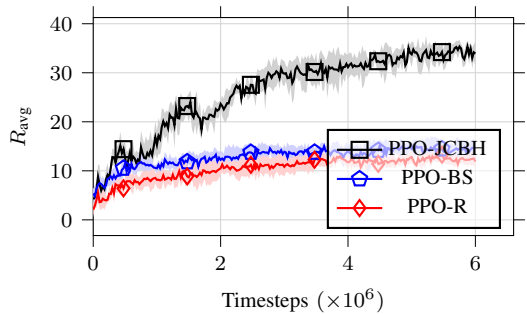
\begin{figure}[!t]
\centering
\begin{tikzpicture}
\begin{axis}[
    width=0.8\columnwidth,
    height=0.5\columnwidth,
    xlabel={Timesteps $(\times10^6)$},
    ylabel={$R_{\text{avg}}$},
    xmin=0,
    axis line style={black},
    tick align=outside,
    tick style={black},
    line width=0.8pt,
    grid=both,
    grid style={draw=gray!15, line width=0.2pt},
    major grid style={draw=gray!35, line width=0.3pt},
    tick label style={font=\footnotesize},
    label style={font=\footnotesize},
    legend pos=south east,
    legend style={
        fill opacity=0.6,
        draw opacity=1.0,
        text opacity=1.0,
        font=\footnotesize
    },
    scaled ticks=true
]

\addplot[name path=ppo_up, draw=none, forget plot]
table[
    col sep=comma,
    x index=0,
    y expr=\thisrow{PPO_H3}+\thisrow{PPO_H3_std}
]{data_HAP/Comparison_1.csv};

\addplot[name path=ppo_low, draw=none, forget plot]
table[
    col sep=comma,
    x index=0,
    y expr=\thisrow{PPO_H3}-\thisrow{PPO_H3_std}
]{data_HAP/Comparison_1.csv};

\addplot[black!18, forget plot]
fill between[of=ppo_up and ppo_low];

\addplot[
    black,
    solid,
    thick,
    mark=square,
    mark size=3.0,
    mark repeat=40,
    mark phase=20
]
table[
    col sep=comma,
    x index=0,
    y=PPO_H3
]{data_HAP/Comparison_1.csv};
\addlegendentry{PPO-JCBH}

\addplot[name path=bs_up, draw=none, forget plot]
table[
    col sep=comma,
    x index=0,
    y expr=\thisrow{Periodic_H3}+\thisrow{Periodic_H3_std}
]{data_HAP/Comparison_1.csv};

\addplot[name path=bs_low, draw=none, forget plot]
table[
    col sep=comma,
    x index=0,
    y expr=\thisrow{Periodic_H3}-\thisrow{Periodic_H3_std}
]{data_HAP/Comparison_1.csv};

\addplot[blue!18, forget plot]
fill between[of=bs_up and bs_low];

\addplot[
    blue,
    solid,
    thick,
    mark=pentagon,
    mark size=3.0,
    mark repeat=40,
    mark phase=20
]
table[
    col sep=comma,
    x index=0,
    y=Periodic_H3
]{data_HAP/Comparison_1.csv};
\addlegendentry{PPO-BS}

\addplot[name path=random_up, draw=none, forget plot]
table[
    col sep=comma,
    x index=0,
    y expr=\thisrow{Random_H3}+\thisrow{Random_H3_std}
]{data_HAP/Comparison_1.csv};

\addplot[name path=random_low, draw=none, forget plot]
table[
    col sep=comma,
    x index=0,
    y expr=\thisrow{Random_H3}-\thisrow{Random_H3_std}
]{data_HAP/Comparison_1.csv};

\addplot[red!18, forget plot]
fill between[of=random_up and random_low];

\addplot[
    red,
    solid,
    thick,
    mark=diamond,
    mark size=3.0,
    mark repeat=40,
    mark phase=20
]
table[
    col sep=comma,
    x index=0,
    y=Random_H3
]{data_HAP/Comparison_1.csv};
\addlegendentry{PPO-R}

\end{axis}
\end{tikzpicture}
\caption{Evolution of learning for JCBH in a stationary environment. Shaded regions indicate $\pm$ one standard deviation.}
\label{fig:comparison_H3_full}
\vspace{-3mm}
\end{figure}
\begin{table}[!t]
\centering
\caption{JCBH in a stationary environment}
\begin{tabular}{lccccc}
\toprule
Policy & $R_{\text{avg}}$ & $\bar{r}_{\text{avg}}$ &   $\bar{s}_{\text{avg}}$ &$\bar{c}_{\text{avg}}$ & $\bar{f}_{\text{srv,avg}}$ \\
\midrule
PPO-JCBH &  31.39 & 19.67 & 14.83\% & 88.91\% & 12.59\%\\
PPO-R & 13.86 & 8.30 & 6.18\% & 73.87\% & 4.98\%  \\
PPO-BS  & 13.82 & 8.44 & 5.38\% & 75.58\% & 4.82\%  \\

\bottomrule
\end{tabular}
\vspace{-3mm}
\label{tab:H3B2}
\end{table}

To evaluate PPO-BH, HAPS are placed above the centres of distinct hotspots, and we compare PPO-BH with random beam (RB) and beam sweeping (BS) for $H = 3$, $B = 2$, and $\tau_2=500$. BS consists of $B$ equidistant beams in terms of azimuth angles and sweeping across the $M$ sub-zones. Experimental results in Table \ref{tab:policy_comparison_BH} demonstrate that PPO-BH consistently outperforms both RB and BS. Given the spatial distribution and transient nature of users, achieving a $100\%$ success rate is unattainable in this environment since the complete geographical area cannot be covered by just three HAPS. PPO-BH achieves $\bar{s}_{\text{avg}} = 20\%$ across all users. Within the coverage region of the HAPS, PPO-BH attains $\bar{c}_{\text{avg}}=95\%$, which represents a $26\%$ improvement over BS.

\begin{table}[!t]
\centering
\caption{JCBH in a periodic environment}
\label{tab:policy_comparison_compact}
\begin{tabular}{lccccc}
\toprule
Policy & $R_{\text{avg}}$ & $\bar{r}_{\text{avg}}$ &   $\bar{s}_{\text{avg}}$ &$\bar{c}_{\text{avg}}$ & $\bar{f}_{\text{srv,avg}}$ \\
\midrule
mPPO-JCBH  & 14.57 & 6.84 & 13.01\% & 79.63\% & 10.57\%\\
mPPO-R & 6.59 & 2.77 & 5.66\% & 53.87\% & 3.98\%  \\
mPPO-BS  & 6.80 & 2.86 & 5.87\% & 54.28\% & 3.85\%  \\

static mPPO-BH  & 3.05 & 1.21 & 2.59\% & 29.86\% & 1.62\%  \\
\bottomrule
\end{tabular}
\label{tab:PeriodicH3}
\vspace{-5mm}
\end{table}

PPO-JCBH is applied to jointly optimize HAPS positioning and BH in a stationary environment with $H=3$ and $B=2$, demonstrating superior performance over heuristic methods. Fig. \ref{fig:comparison_H3_full} compares learning for PPO-JCBH, PPO-R, and PPO-BS. PPO-R uses PPO for positioning with random BH, while PPO-BS employs PPO for positioning with the reward defined in \eqref{eq:rewardBH} and uses beam sweeping for BH. We notice that PPO-BS and PPO-R converge faster than PPO-BH, as they have a lower action space. However, PPO-JCBH converges to a three times higher reward in comparison to PPO-R and PPO-BS, showing that just positioning is not sufficient, dynamic adapted beaming strategy makes the difference to allow relevant service from the HAPS. In Table \ref{tab:H3B2}, PPO-JCBH outperforms PPO-BS and PPO-R in terms of $\bar{r}_{\text{avg}}$ and $\bar{s}_{\text{avg}}$, demonstrating that learning a BH strategy enables more effective allocation in scenarios with uneven user density.

Building on these positive results in the stationary environment, which demonstrate the feasibility of simultaneously learning positioning and beam allocation, the study is further extended to periodic environments. Table \ref{tab:PeriodicH3} compares a multi-policy mPPO-JCBH solution with mPPO-R and mPPO-BS, which are multi-policy PPO solutions for HAPS positioning with random beam and beam sweeping, respectively. We also compare with static HAPS with mPPO for BH, where mPPO-JCBH outperforms all other solutions. However, in the periodic environment, both $\bar{s}_{\text{avg}}$ and $\bar{r}_{\text{avg}}$ are reduced compared to the stationary case due to HAPS traversing low-density areas during phase transitions. Notably, static mPPO-BH achieves significantly lower performance, underscoring the importance of HAPS mobility. These results confirm that RL-based JCBH provides substantial performance gains over separate or static strategies.

\section{Conclusion}
In this work, we have proposed a centralized RL framework for joint coverage and beam optimization in multi-HAPS systems under non-stationary traffic conditions. A spatio-temporal user model is considered to capture realistic dynamics. PPO is first applied to HAPS positioning, where a multi-policy approach improves adaptation to periodic traffic patterns. Next, a PPO-based beam-hopping strategy is developed, outperforming heuristic methods such as random allocation and beam sweeping in terms of sum-rate and user satisfaction, including average success. A unified RL framework is then proposed for joint optimization. The joint approach consistently outperforms decoupled solutions, achieving more than a twofold improvement in average aggregate success, along with notable gains in sum-rate. These results highlight the potential of RL to exploit HAPS mobility for dynamic, demand-aware network adaptation. To the best of our knowledge, this is among the first works to jointly optimize HAPS positioning and beam-hopping under non-stationary traffic with explicit demand satisfaction objectives. Future work will consider decentralized multi-agent RL and more realistic atmospheric conditions.

\bibliographystyle{IEEEtran}
\bibliography{library.bib}

\end{document}